\documentclass{mn2e}
\usepackage{epsf,times}
\newcommand{\etal}{{et al}\/.}
\begin{document}
\title[Unifying B2 radio galaxies with BL Lac objects]{Unifying B2
radio galaxies with BL Lac objects}
\author[M.J.~Hardcastle \etal]{M.J.\ Hardcastle$^1$\thanks{E-mail: m.hardcastle@bristol.ac.uk.},
D.M.\ Worrall$^{1,2}$, M.\ Birkinshaw$^{1,2}$ and C.M.\
Canosa$^1$\thanks{Present address: British Telecommunications plc, PP
D007, North Star House, North Star Avenue, Swindon, Wiltshire, SN2 1BS}\\
$^1$ Department of Physics, University of Bristol, Tyndall Avenue,
Bristol BS8 1TL\\
$^2$ Harvard-Smithsonian Center for Astrophysics, 60 Garden Street,
Cambridge, MA 02138, U.S.A.}
\maketitle
\begin{abstract}
In an earlier paper we presented nuclear X-ray flux densities,
measured with {\it ROSAT}, for the B2 bright sample of nearby
low-luminosity radio galaxies. In this paper we construct a nuclear
X-ray luminosity function for the B2 radio galaxies, and discuss the
consequences of our results for models in which such radio galaxies
are the parent population of BL Lac objects. Based on our observations
of the B2 sample, we use Monte Carlo techniques to simulate samples of
beamed radio galaxies, and use the selection criteria of existing
samples of BL Lac objects to compare our simulated results to what is
observed. We find that previous analytical results are not applicable
since the BL Lac samples are selected on beamed flux density. A simple
model in which BL Lacs are the moderately beamed ($\gamma \sim 3$)
counterparts of radio galaxies, with some random dispersion ($\sim
0.4$ decades) in the intrinsic radio-X-ray relationship, can reproduce
many of the features of the radio-selected and X-ray-selected BL Lac
samples, including their radio and X-ray luminosity functions and the
distributions of their radio-to-X-ray spectral indices. In contrast,
models in which the X-ray and radio emission have systematically
different beaming parameters cannot reproduce important features of
the radio-galaxy and BL Lac populations, and recently proposed models
in which the radio-to-X-ray spectral index is a function of source
luminosity cannot in themselves account for the differences in the
slopes of the radio and X-ray-selected BL Lac luminosity functions.
The redshift distribution and number counts of the X-ray-selected EMSS
sample are well reproduced by our best models, supporting a picture in
which these objects are beamed FRI radio galaxies with intrinsic
luminosities similar to those of the B2 sample. However, we cannot
match the redshift distribution of the radio-selected 1-Jy sample, and
it is likely that a population of FRII radio galaxies is responsible
for the high-redshift objects in this sample, in agreement with
previously reported results on the sample's radio and
optical-emission-line properties.
\end{abstract}

\section{Introduction}
\label{intro}
The extreme properties of BL Lac objects are explained in terms of
relativistic beaming of the emission from a jet oriented close to the
line of sight. This model implies the existence of a substantial
`parent population' of sources whose jets are less favourably aligned,
and it is widely accepted that this is the population of low-power
radio galaxies (Browne 1983; Urry \& Padovani 1995). Properties which
are isotropic and unaffected by beaming should be similar in BL Lac
objects and low-power radio galaxies. This is broadly supported by
observations of extended radio emission (e.g.,\ Antonucci \& Ulvestad
1985, Kollgaard \etal\ 1992, Perlman \& Stocke 1993, 1994), host
galaxies (e.g., Ulrich 1989, Abraham, McHardy \& Crawford 1991; Wurtz,
Stocke \& Yee 1996; Falomo 1996) and cluster environments (Pesce,
Falomo \& Treves 1995; Smith, O'Dea \& Baum 1995; Wurtz, Stocke \&
Ellingson 1997).

In a series of papers (Padovani \& Urry 1990; Padovani \& Urry 1991;
Urry, Padovani \& Stickel 1991), Urry and co-workers made this model
quantitative by predicting the luminosity function of BL Lac objects
based on that of radio galaxies. They used the analysis of Urry \&
Schafer (1984) and Urry \& Padovani (1991) to calculate the expected
luminosity function of a population of beamed objects given a parent
(unbeamed) luminosity function. With the data on radio-galaxy and BL
Lac populations then available they were able to show reasonable
agreement between the predictions of the model and the observed
luminosity functions. They found that the luminosity function of
radio-selected BL Lac objects from the 1-Jy sample (Stickel \etal\
1991) was consistent with that of radio galaxies from the 2-Jy sample
(Wall \& Peacock 1985) if the radio-emitting plasma in the cores had a
bulk Lorentz factor $\gamma_{\rm radio} > 5$. The X-ray
luminosity function and number counts of X-ray selected BL Lac objects
were consistent with the luminosity function of FRI radio galaxies
observed with {\it Einstein} by Fabbiano \etal\ (1984) if the X-ray
emitting plasma has a somewhat lower bulk Lorentz factor, $\gamma_{\rm
Xray} \approx 3$.

The issue of the relationship between radio galaxies and X-ray and
radio-selected BL Lac objects is only partially resolved by this work.
The best-studied sample of X-ray selected BL Lacs is the EMSS sample
(Wolter \etal\ 1991, Rector \etal\ 2000) and the best radio-selected
sample is the 1-Jy sample (Stickel \etal\ 1991, Rector \& Stocke
2001). When these samples are compared, a number of differences
emerge. Some -- perhaps as many as half -- of the 1-Jy objects have
radio structures, luminosities and emission-line properties similar to
those of FRII radio galaxies (Antonucci \& Ulvestad 1984, Kollgaard \etal\
1992, Rector \& Stocke 2001) while the EMSS BL Lacs are all FRI-like
(Perlman \& Stocke 1993). The luminosity functions of the two samples
are also different; the radio-selected objects show evidence for
positive evolution (i.e. sources were more numerous or more powerful
at higher redshift) while X-ray-selected sources appear to be {\it
negatively} evolving. Urry \etal\ were forced by the data then
available to use an X-ray-selected sample of BL Lacs for their
comparison with the X-ray luminosity function of radio galaxies, and a
radio-selected sample when considering the radio luminosity function.
They were thus unable to say anything about the relation between radio
galaxies and the two BL Lac populations.

Radio-selected BL Lacs are often considered to be more extreme than
X-ray selected objects; they have more prominent radio cores (Perlman
\& Stocke 1993, Laurent-Muehleisen \etal\ 1993, Rector \& Stocke
2001), higher optical polarization with less stable position angle
(Jannuzi, Smith \& Elston 1994) and a lower optical starlight fraction
(Stocke \etal\ 1985). On the other hand, their environments and host
galaxies are similar to X-ray selected sources (Wurtz, Stocke \& Yee
1996, Wurtz \etal\ 1997). This has led to suggestions that
radio-selected objects are more strongly affected by Doppler boosting
than X-ray selected objects, and so are being observed at smaller
angles to the line of sight. Since radio-selected and X-ray selected
objects have similar X-ray luminosities, this requires that the
X-ray-emitting regions be less strongly beamed, or even isotropic
(Maraschi \etal\ 1986; Celotti \etal\ 1993); the idea of weaker
beaming is consistent with the difference in the Doppler beaming
factors found by Urry \etal\ (1991). This in turn implies that
X-ray-selected BL Lacs should be the more numerous population, since
they can be seen at larger angles to the line of sight, which Celotti
\etal\ (1993) argue is consistent with the X-ray luminosity functions
of the two populations. Models of this kind can either rely on
differences in velocity between X-ray and radio-emitting regions
(e.g., Ghisellini \& Maraschi 1989) or differences in opening angle of
the jet (Celotti \etal\ 1993). However, these models do not account
for the differences in evolutionary properties between the two
samples. In section \ref{beaming} of this paper we shall discuss
whether they are consistent with observations of radio galaxies and
with the X-ray and radio luminosity functions of the two classes.

A description of BL Lac objects in terms of the selection waveband
does not necessarily reflect the underlying physics. An alternative
approach is to refer to high-energy peaked BL Lacs (HBL) and
low-energy peaked BL Lacs (LBL) (e.g., Giommi \& Padovani 1994,
Padovani \& Giommi 1995), distinguishing between the two classes by
their radio/X-ray flux ratios; a typical dividing line is that HBL
have $\log_{10}(F_{\rm 1\ kev}/F_{\rm 5\ GHz}) >-5.5$, or,
equivalently, $\alpha_{\rm RX}<0.72$, where $\alpha$ is defined here
and throughout the paper in the sense $F \propto \nu^{-\alpha}$ and
$\alpha_{\rm RX}$ denotes the radio-to-X-ray two-point spectral index. By
selecting bright sources at one or the other waveband we may simply be
picking up objects with extreme radio/X-ray flux ratios, which
suggests various possible schemes for unifying the two populations
(e.g., Giommi \& Padovani 1994, Fossati \etal\ 1997) and is consistent
with the discovery of `intermediate' BL Lacs in deeper surveys
(Laurent-Muehleisen \etal\ 1999). From multi-wavelength observations
it has been found that the spectral energy distributions (SEDs) of BL
Lacs can be represented by a low-frequency peak (assumed to be
synchrotron radiation) and a high-frequency peak (perhaps
inverse-Compton radiation). There is evidence that the positions of
these peaks shift to higher frequencies at lower bolometric
luminosities (Sambruna, Maraschi \& Urry 1996, Fossati \etal\ 1998).
However, Giommi, Menna \& Padovani (1999) do not find the increase in
the numbers of HBL at fainter radio fluxes that is expected in such a
model.

To try to resolve some of the outstanding issues in BL Lac unification
it is productive to learn more about the presumed parent population of
many or all of the BL Lac objects, namely the low-power radio
galaxies. Work to date has been hampered by the lack of a
well-defined, low-frequency-selected sample of radio galaxies with
well-known radio and X-ray properties. In a previous paper (Canosa
\etal\ 1999) we presented results for the 40 members ($\approx 80$ per
cent) of the B2 bright sample of radio galaxies which had been
observed in pointed {\it ROSAT} observations. The high spatial
resolution of {\it ROSAT} allowed us to separate the nuclear emission
from extended emission due to the hot-gas environment of the radio
source. In this paper we use these data to derive a new X-ray
luminosity function for the nuclei of B2 radio sources, largely free
from contamination from thermal emission from the sources' hot-gas
environments. We use numerical techniques to extend earlier work,
first asking how well a simple beaming model can be used to link the
X-ray and radio luminosity functions of radio galaxies, and then
investigating the expected relationship between radio galaxies and BL
Lac objects when selection bias is taken into account.

Throughout the paper we use a cosmology in which $H_0 = 50{\rm\
km\,s^{-1}\,Mpc^{-1}}$ and $q_0 = 0$.

\section{Luminosity functions of B2 bright galaxies}
\label{b2lum}

The B2 bright sample, described by Colla \etal\ (1975), consists of 52
radio sources with 408-MHz radio flux $\ga 0.20$ Jy identified with
galaxies with photographic magnitudes $\la 15.6$. (The flux limit is
declination-dependent; see Colla \etal\ for details.) The sample is
flux-complete for all but the largest sources. The sky coverage of the
sample is 1.154 sr [cf.\ Colla (1975); the figure quoted by Ulrich
(1989) appears to be in error]. Of the 52 objects in the sample, two
(B2 0207+38 and 1318+34) are starburst galaxies, and will not be
considered further in this paper. One (B2 1833+32) is the broad-line
FRII radio galaxy 3C\,382, which we also exclude from further
consideration because FRII radio sources do not participate in most
radio galaxy-BL Lac unification models, and two (B2 1101+38 and
1652+39) are the well-known BL Lac objects Mrk 421 and Mrk 501, which
we did not discuss in Canosa \etal\ (1999), but which will be included
in the sample we discuss here in order that it should be unbiased with
respect to orientation\footnote{We have obtained X-ray flux densities
for these objects from archival {\it ROSAT} data, in order to match
our analysis of the radio galaxies. For Mrk 421 there were four
observations at different epochs in 1992--3, and the flux density we
use corresponds to the median count rate in these observations.}. The
remainder are more or less typical FRI radio sources, though the
sample includes some peculiar objects such as B2 1626+39 (3C\,338).
Ulrich (1989) has shown that the B2 bright sample is well matched in
{\it unbeamed} properties, like extended radio luminosity, to samples
of nearby BL Lac objects.

\begin{table*}
\caption{The sample of B2 radio galaxies and their properties.}
\label{sample}
\begin{tabular}{llrrrrrr}
\hline
B2 name&Other name&$z$&$S_{408}$&$\alpha_{408}$&$S_{\rm 5,\
core}$&Reference&$S_{\rm 1\ keV,\ core}$\\
&&&(Jy)&&(mJy)&&(nJy)\\
\hline
0034+25 & &0.0321 &0.29 &0.63 &10 &1 &-- \\
0055+30 &NGC 315 &0.0167 &3.01 &0.33 &450 &3 &$190 \pm10 $\\
0055+26 &NGC 326 &0.0472 &5.26 &0.81 &8.6 &4 &$11 \pm2 $\\
0104+32 &3C\,31 &0.0169 &10.6 &0.62 &92 &1 &$64 \pm7 $\\
0116+31 &4C\,31.04 &0.0592 &3.75 &0.39 &$<$100 &7 &-- \\
0120+33 &NGC 507 &0.0164 &0.74 &1.4 &1.4 &1 &$<$83 \\
0149+35 &NGC 708&0.0160 &0.39 &0.6 &5 &1 &$<$59 \\
0206+35 &4C\,35.03 &0.0375 &4.90 &0.63 &106 &1 &$29 \pm5 $\\
0222+36 & &0.0327 &0.37 &0.21 &90 &1 &$<$10 \\
0258+35 &NGC 1167 &0.016 &4.19 &0.51 &$<$15 &1 &$<$10 \\
0326+39 & &0.0243 &2.04 &0.53 &70 &1 &$36 \pm4 $\\
0331+39 &4C\,39.12 &0.0202 &1.85 &0.49 &125 &1 &$230 \pm20 $\\
0722+30 & &0.0191 &0.41 &1.08 &51 &1 &$<$20 \\
0755+37 &NGC 2484 &0.0413 &6.34 &0.65 &190 &1 &$85 \pm9 $\\
0800+24 & &0.0433 &0.34 &0.65 &3 &1 &$<$3 \\
0836+29A &4C 29.30 &0.0650 &1.68 &0.62 &8.2 &1 &$<$4 \\
0844+31 &4C 31.32 &0.0675 &4.03 &0.75 &58 &1 &-- \\
0915+32 & &0.0620 &0.50 &0.45 &8 &1 &-- \\
0924+30 & &0.0266 &1.88 &1.01 &$<$0.4 &1 &$<$20 \\
1040+31 &4C 29.41 &0.036 &1.82 &0.59 &55 &1 &$16 \pm2 $\\
1101+38 &Mrk 421 &0.03 &1.24 &0.23 &520 &8 &$7.0 \times 10^4 $\\
1102+30 & &0.0720 &1.0 &0.69 &26 &1 &-- \\
1108+27 &NGC 3563 &0.0331 &0.21 &0.45 &14 &2 &-- \\
1113+29 &4C\,29.41 &0.0489 &5.32 &0.53 &41 &1 &$10 \pm6 $\\
1122+39 &NGC 3665 &0.0067 &0.31 &0.61 &6 &1 &$1 \pm2 $\\
1217+29 &NGC 4278 &0.0021 &0.62 &0.21 &350 &1 &$100 \pm20 $\\
1254+27 &NGC 4839 &0.02464 &0.23 &0.83 &2.3 &2 &$25 \pm6 $\\
1256+28 &NGC 4869 &0.0224 &1.57 &1.01 &2 &1 &$<$2 \\
1257+28 &NGC 4874 &0.0239 &0.54 &0.73 &1.1 &1 &$6 \pm2 $\\
1317+33 &NGC 5098 &0.0379 &0.26 &0.90 &8 &6 &$7 \pm4 $\\
1321+31 &NGC 5127 &0.0161 &3.87 &0.62 &21 &1 &$<$10 \\
1322+36 &NGC 5141 &0.0175 &1.84 &0.43 &75 &1 &$<$10 \\
1346+26 &A 1795 &0.0633 &3.38 &0.89 &53 &1 &$36 \pm10 $\\
1350+31 &3C\,293 &0.0452 &10.90 &0.67 &100 &1 &$<$20 \\
1422+26 & &0.0370 &2.17 &0.71 &25 &1 &$4 \pm1 $\\
1525+29 &A 2079 &0.0653 &0.46 &0.70 &2.5 &1 &-- \\
1553+24 & &0.0426 &0.21 &0.25 &53.6 &2 &$28 \pm6 $\\
1610+29 &NGC 6086 &0.0313 &0.34 &0.69 &$<$6 &1 &$12 \pm3 $\\
1615+35 &NGC 6109 &0.0296 &5.80 &0.79 &28 &1 &$13 \pm2 $\\
1621+38 &NGC 6137 &0.031 &1.00 &0.53 &50 &1 &$16 \pm3 $\\
1626+39 &3C\,338 &0.0303 &19.46 &1.32 &100 &1 &$20 \pm7 $\\
1652+39 &Mrk 501 &0.034 &1.90 &0.15 &1420 &5 &$1.7 \times 10^4  $\\
1855+37 & &0.0552 &0.9 &0.81 &$<$100 &1 &$35 \pm6 $\\
2116+26 & &0.0164 &0.34 &0.02 &47 &1 &-- \\
2229+39 &3C\,449 &0.0171 &6.93 &0.61 &37 &1 &$15 \pm5 $\\
2236+35 & &0.0277 &0.83 &0.55 &8 &1 &$14 \pm4 $\\
2335+26 &3C\,465 &0.0301 &21.7 &0.75 &230 &1 &$66 \pm6 $\\
\hline
\end{tabular}
\vskip6pt
\begin{minipage}{11.5cm}
408-MHz flux densities and $\alpha_{\rm 408}$ (the low-frequency
spectral index) have been corrected to the scale of Baars \etal\
(1977) as described in the text. 5-GHz flux densities for the cores
are taken from the literature. References for these are as follows:
(1) Giovannini \etal\ (1988) (2) Fanti \etal\ (1987) (from 1.4 GHz)
(3) Venturi \etal\ (1993) (4) Fomalont, private communication (5)
Stickel \etal\ (1991) (6) Measured from maps supplied by R. Morganti
(7) Cotton \etal\ (1995) (8) Antonucci \& Ulvestad (1985)
(extrapolated from 1.5 GHz). 1-keV nuclear flux densities are taken
from Canosa \etal\ (1999), except for those for Mrk 421 and 501, which
are discussed in the text, and for B2 1553+24, for which Canosa \etal\
quoted an incorrect X-ray flux density, although the count rate they
quote is correct.
\end{minipage}
\end{table*}

Because an unbiased selection on unbeamed properties is important, we
need to consider whether some of the sample sources exceed the 408-MHz
flux limit solely because of a beamed radio component. We may estimate
this by subtracting the 5-GHz core flux density from the total 408-MHz
flux, on the assumption that the core spectrum is flat between these two
frequencies, and seeing whether the result lies below the B2
position-dependent flux limit. On this basis, 4 sources (B2 0648+27,
1108+27, 1144+35 and 1553+24) possibly have too low a 408-MHz flux to
be in the sample through extended emission alone. Of these, the core
of 1144+35 is known to dominate the extended structure, and to be
variable (Giovannini \etal\ 1999). 0648+27 is also core-dominated
(e.g., Antonucci 1985). We exclude these two sources from the sample.
The other two are reasonably normal members of the sample which happen
to lie at the flux limit, so that the case for excluding them (given
the large uncertainties in core spectrum) is uncertain -- Laing \etal\
(1999), on a similar basis, chose not to exclude these two sources
from their sample of jetted objects. Our revised, unbiased sample then
consists of 47 objects, whose properties are listed in Table
\ref{sample}. We have nuclear X-ray information for all but 8 of
these.

\begin{figure*}
\epsfxsize 10cm
\epsfbox{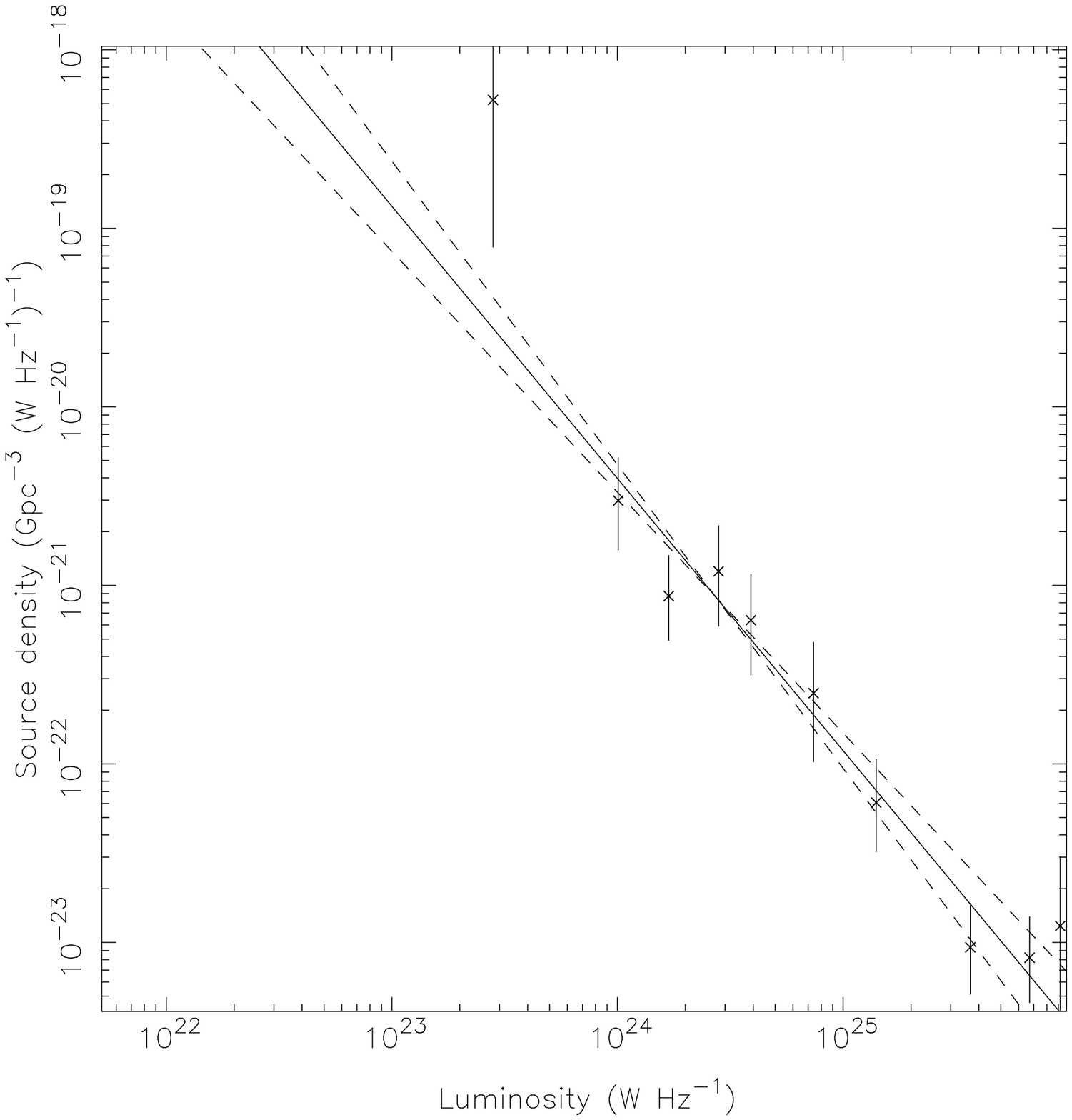}
\caption{Differential 408-MHz luminosity function for extended flux from the B2
sample. The solid line shows the best-fit linear regression slope to
the data, $-1.52 \pm 0.13$, while the dashed lines give an indication
of the $1\sigma$ error. Each bin represents 5 sources.}
\label{408-lf}
\end{figure*}

We are now in a position to construct a luminosity function for the
extended radio emission from our unbiased sample. For a source which
has a detected radio core at 5 GHz, we subtract its flux density from
the 408-MHz flux density before making the luminosity calculation; for
sources without detected cores (where the upper limit on core flux
density is typically a small fraction of the total flux density) we
just use the total 408-MHz flux density, and we also do this for
1108+27 and 1553+24. For the purposes of calculating luminosities we
correct the 408-MHz flux densities to the Baars \etal\ (1977) scale
from the Kellerman (1964) scale used in B2 the catalogue (Colla \etal\
1970), using the factor 1.074 suggested by Baars \etal . Spectral
indices are also corrected to the Baars \etal\ scale in the same way.
To generate a differential luminosity function we use the $1/V_{\rm
max}$ method, taking into account both the optical and radio selection
criteria; the optical selection criterion is the important one in
31/47 objects. The radio flux limits and flux densities used to
determine $V_{\rm max}$ are the original values, without core
subtraction. We assign errors to the bins in the luminosity function
using the method of Wolter \etal\ (1994), having verified by
simulation that the error estimates derived from this method are close
to the real statistical errors. Our differential luminosity function
for the extended flux from the sample is plotted in Fig. \ref{408-lf}.
It is consistent with the luminosity function derived for the sample
by de Ruiter \etal\ (1990).

We can similarly derive 5-GHz and 1-keV luminosity functions for the
nuclei of the sample, using the method described by Wolter \etal\
(1994). The values of $V_{\rm max}$ (or $V_{\rm a}$ in their notation)
are derived from the selection band (i.e. the 408-MHz flux densities
and optical magnitudes) while the luminosities are those in the
observation band. The 5-GHz radio core flux densities include 5 upper
limits; this is a sufficiently small fraction of the data that we are
not going to be grossly in error if we assume that the true values of
the core fluxes for these sources are equal to the limits (i.e. treat
limits as detections at the limit level). Doing this leads to the
luminosity function plotted in Fig. \ref{5-lf}. (Throughout the paper
we assume that the radio spectral index $\alpha$ of the cores is 0.)

However, the 1-keV flux densities include 12 upper limits and 8
unobserved sources. In most cases the upper limits are the result of
an ambiguity in the separation of nuclear and extended X-ray emission.
Since we have demonstrated that the unobserved sources have no known
bias with respect to the rest of the sample (Canosa \etal\ 1999), we
can simply scale the luminosity function computed for the other
sources by a factor 47/39. But the large number of X-ray limits
(nearly one third of the total observed sources), could seriously
affect the derived luminosity function. To assess the effect of the
limits, we computed the luminosity function in four different ways:
(a) with the limits treated as detections; (b) with the non-detected
sources omitted altogether and the results scaled; (c) with the limits
treated as detections at 0.1 times the limit level; and (d) with the
limits replaced by the values that the X-ray cores `should' have
according to the radio-X-ray core correlation reported in Canosa
\etal\ (1999). The resulting luminosity functions were similar, with
slopes and normalizations of straight-line fits to the data being
consistent within the statistical uncertainties, so our results here
are insensitive to the treatment of the upper limits. Accordingly, we
treat the limits as detections in order to generate Fig.\ \ref{1-lf},
which has also been scaled to take account of the unobserved sources.
(Throughout the paper we assume that the X-ray spectral index of cores
is $\alpha = 0.8$.) The result is the first nuclear X-ray luminosity
function to be computed for a sample of radio galaxies.

Since the analysis of Canosa \etal\ (1999), we have obtained the first
{\it Chandra} images of FRI radio galaxies from the B2 sample (Worrall
\etal\ 2001; Hardcastle \etal\ 2002). These show that in some of
these objects there is extended X-ray emission on scales of a few
arcseconds, in addition to a point-like nucleus and the larger-scale
group or cluster emission that we have discussed elsewhere (Canosa
\etal\ 1999; Worrall \& Birkinshaw 2000). This small-scale extended
emission would have been unresolved to {\it ROSAT}, and therefore
contaminates the nuclear 1-keV flux densities we quote in this paper.
But it does not dominate the total soft-X-ray flux in the B2 objects we
have observed with {\it Chandra} so far, and we know from the
correlation between radio, X-ray and optical nuclear emission in these
objects (Canosa \etal\ 1999; Hardcastle \& Worrall 1999, 2000) that a
significant fraction of the {\it ROSAT}-observed 1-keV emission must
arise in a beamed nucleus. It is therefore reasonable to use the
current values to test the predictions of unified models, provided we
bear in mind the possible distorting effects of this contaminating
extended emission.
\label{chandra-caveat}

\section{Luminosity functions for BL Lac objects}
\label{bllum}

To test unified models using the new B2 luminosity function we must
have comparison samples of BL Lac objects. The two comparison samples
we shall consider in this paper are the 1-Jy sample of radio-selected
BL Lacs [we use the 34-object sample of Stickel \etal\ (1991), because
of the good availability of X-ray information, but with the redshifts
tabulated by Rector \& Stocke (2001)] and the EMSS sample of X-ray
selected objects [we use the 41-object `D40' sample of Rector \etal\
(2000)]. Flux densities are now available for almost all objects in
both these samples at both 5-GHz radio and X-ray frequencies (Urry
\etal\ 1996, Fossati \etal\ 1998 and references therein; Rector \etal\
and references therein). So, unlike Urry and co-workers, we are not
limited to comparing the radio luminosity function of the B2s to the
radio luminosity function of radio-selected BL Lacs, and so on; we can
compare the luminosity functions of both kinds of BL Lac in both
wavebands to the radio-galaxy observations.

Accordingly, we have constructed radio and X-ray luminosity functions
for the two samples in the same way as for the radio-galaxy data. In
both cases, we calculate $V_{\rm max}$ using the energy band in which
the source was selected. We take account of the optical flux selection
criterion for the 1-Jy sample ($<20$ mag on sky survey plates;
Stickel \etal 1991) and the variable sensitivity of the EMSS (Gioia \etal\
1990) in the manner described by Morris \etal\ (1991). In both cases,
we include modest luminosity evolution as an exponential function of
look-back time (positive for the radio-selected objects, negative for
the X-ray-selected objects), in the manner described by Wolter \etal\
(1994), to obtain $<V/V_{\rm max}> \approx 0.5$;
the effect of this assumption on the results is quite small. For the
sources with only lower redshift limits, we have used the limit as
though it were the actual redshift of the source. For the few sources
without known redshifts (5 in the 1-Jy sample, 2 in the EMSS sample),
we have used the median redshift for the detected objects in the
sample from which they were drawn (0.55 for the 1-Jy sample, 0.30 for
the EMSS sample). We have constructed luminosity functions on the
assumption that the nuclear components in both radio and X-ray are
dominant in both classes of source.

\begin{figure*}
\epsfxsize 10cm
\epsfbox{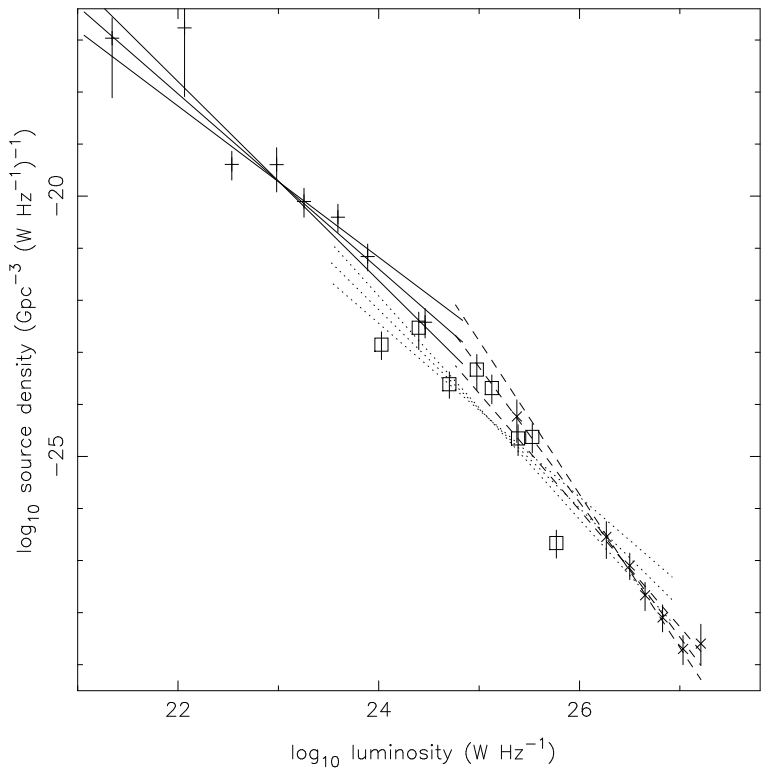}
\caption{Differential 5-GHz core luminosity function for the B2 sample and
the comparison samples of BL Lac objects. The inner lines show the
best-fit linear regression slope to the data, while the two outer
lines show the $1\sigma$ error. Crosses mark data points from the B2
sample, Xs the 1-Jy sample and boxes the EMSS sample, while solid
lines denote fits to the B2s, dashed lines the 1-Jy sample and dotted
lines the EMSS sample. The slope of the luminosity function for the B2
sample is $-1.68 \pm 0.16$, for the 1-Jy sample $-2.59 \pm 0.26$, and
for the EMSS sample $-1.91 \pm 0.18$. Each bin represents 5 sources.}
\label{5-lf}
\label{5-lfbl}
\end{figure*}

\begin{figure*}
\epsfxsize 10cm
\epsfbox{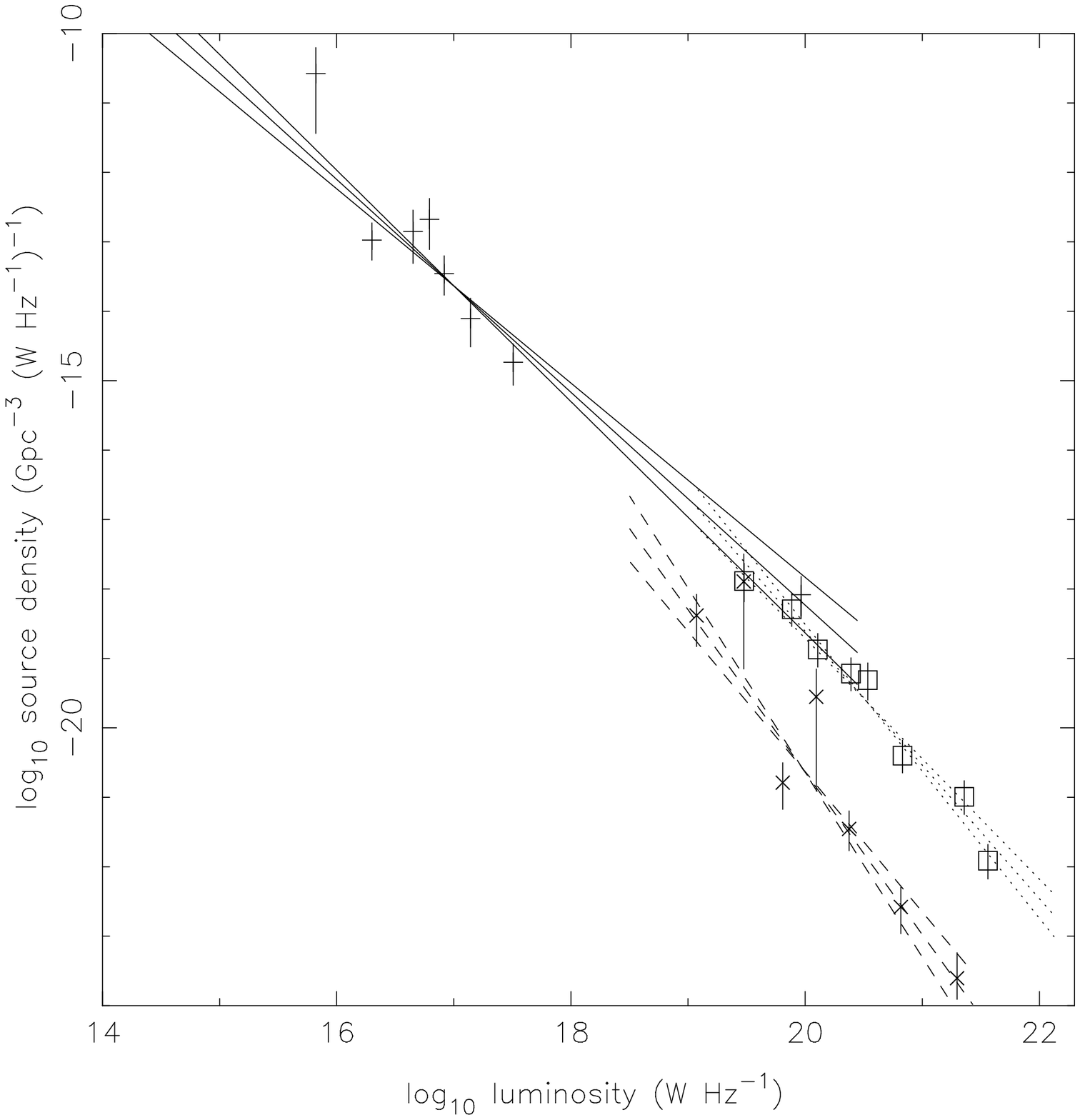}
\caption{Differential 1-keV nuclear luminosity function for the B2
sample and the BL Lac objects. Lines and symbols are as in Fig.\
\ref{5-lf}. The slope of the luminosity function for the B2 sample is
$-1.53 \pm 0.09$, for the 1-Jy sample $-2.34 \pm 0.21$, and for the
EMSS sample $-1.93 \pm 0.13$. Each bin represents 5 sources.}
\label{1-lf}
\label{1-lfbl}
\end{figure*}

The results are plotted in Figs \ref{5-lfbl} and \ref{1-lfbl}. They
are, as we would expect, consistent on the whole with other
determinations of the luminosity functions for these samples
(e.g., Stickel \etal\ 1991, Wolter \etal\ 1994). 


\section{Application of the Urry \& Schafer model}
\label{usmodl}

Urry \& Schafer (1984) discuss the effect of relativistic beaming on a
differential luminosity function which is intrinsically a simple power
law (i.e., one which has the form $\phi(L) = K L^{s}$, where $L$ is
the luminosity, $\phi$ has dimensions volume$^{-1}$ luminosity$^{-1}$,
and $s$ is the slope of the luminosity function.) They show that the
observed luminosity function is a broken power law. The low-luminosity
slope is given by $-(p+1)/p$, where $p$ is the power to which the
Doppler factor is raised in the beaming calculation; for a jet, $p = 2
+ \alpha$, where $\alpha$ is the jet spectral index. At high
luminosities, the slope of the observed luminosity function is the
same as that of the intrinsic luminosity function.

The luminosity functions for the radio and X-ray cores of the B2
objects shown in Figs \ref{5-lf} and \ref{1-lf} are for beamed
components of the radio source. The unbeamed 408-MHz radio luminosity
function in Fig.\ \ref{408-lf} can be treated as a single power law,
since the well-known steepening (e.g., Machalski \& Godlowski 2000) in
the low-frequency radio source luminosity function occurs around
$L_{\rm 408 MHz} \ga 4 \times 10^{25}$ W Hz$^{-1}$. If we make the
further assumption that the {\it intrinsic} core radio and X-ray
fluxes of the B2s are proportional to the extended 408-MHz flux, the
Urry \& Schafer analysis can be applied to these data. Since most
sources are at large angles to the line of sight, we are predominantly
seeing the low-luminosity end of the beamed luminosity function. Urry
\& Schafer's result leads us to expect slopes of $-1.50$ and $-1.36$
for the 5-GHz and 1-keV core luminosity functions, respectively. The
actual slopes ($-1.67 \pm 0.15$ and $-1.54 \pm 0.09$ respectively) are
somewhat steeper than, but consistent with, these expectations,
supporting our assumption above about the relationship between total
luminosity and intrinsic core luminosity. Since the slope of the
unbeamed, `parent' luminosity function should be equal to that of the
408-MHz data, $-1.52 \pm 0.13$, which is similar to the predicted
slope of the beamed luminosity function, we would not expect to see a
break in the slope of the beamed luminosity function at high
luminosities, and indeed there is no evidence for such a break in the
luminosity functions for the beamed emission from the B2s.

What would we expect to see on this model for the BL Lac samples if
the B2 sources were their parent population? The BL Lac objects are
clearly at the luminous end of the beamed luminosity function, and
therefore we should recover the slope of the parent luminosity
function. Since this is $\sim 1.5$, a simple application of the Urry
\& Schafer model would cause us to expect the slopes of the
differential luminosity functions of BL Lac objects that are beamed
versions of the B2 radio galaxies to be $\sim 1.5$ (this result is
independent of the value of $p$). The observed slopes are much steeper
than this, corresponding to a parent population with an unbeamed slope
$\sim 2$. Since the slope of the luminosity function of
higher-luminosity radio galaxies is more similar to 2, we might be
able to account for this by saying that these BL Lacs are beamed
versions of more luminous sources than the B2s. However, the extended
radio luminosities of the BL Lacs are mostly FRI-like, and the
luminosity functions in radio and X-ray of the two populations
overlap. Either the B2s, in spite of these similarities, are not the
parent population of these objects, or the analysis of Urry \& Schafer
is not applicable.

\section{Luminosity functions from simulated observations}

Urry \& Schafer predicted the luminosity functions of beamed populations
using simple mathematical arguments. An alternative approach is to use
Monte Carlo simulations to generate observations of such sources,
based on the information we have on the parent population, and to
generate luminosity functions from this simulated population. In this
section we adopt that approach.

\subsection{Models}
\label{mc}

The initial model we adopt is a simple one. Each Monte-Carlo source is
assigned a low-frequency luminosity drawn from the power-law fit to
the 408-MHz luminosity function of the B2 radio galaxies (Fig.\
\ref{408-lf}), assuming that luminosities lie between $10^{23}$ and $8
\times 10^{25}$ W Hz$^{-1}$. This luminosity range matches the B2
observations, and its lower bound is close to the lower limit on the
luminosity of AGN found when the radio-galaxy luminosity function is
constructed for large samples (Condon 1989; Machalski \& Godlowski
2000). However, it is not clear whether the lack of radio galaxies
below this luminosity indicates a real physical lower limit on
radio-galaxy luminosity, or whether it simply becomes impossible to
distinguish them from the much more numerous starbursts; for example,
Sadler \etal\ (2001), using an optical approach to classification,
find no sign of a turnover, with a luminosity function going as low as
a few $\times 10^{22}$ W Hz$^{-1}$ at 408 MHz. We comment below
(section \ref{matching}) on the effects on unified schemes of allowing
the radio-galaxy luminosity function to extend to lower luminosities.

The rest-frame luminosity of the core in both radio
and X-rays is assumed to be a fixed fraction $f$ of the total
low-frequency luminosity; the observed core luminosity depends on $f$
and on the beaming Lorentz factor and the (randomly chosen) angle to
the line of sight. This implies that the ratio of core luminosity to
extended luminosity, known as the `core prominence', for a given
source is a function only of the Doppler factor ${\cal D} = [\gamma(1
- \beta \cos \theta)]^{-1}$, where $\gamma$ is the Lorentz factor,
$\beta = (1-1/\gamma^2)^{1/2}$, and $\theta$ is the angle to the line of
sight and of the `intrinsic prominence', the core prominence in its
rest frame. This simple model has been used previously to constrain
core Lorentz factors (e.g., Morganti \etal\ 1995; Hardcastle \etal\
1999). Later we will consider the effects of introducing some
dispersion into the relation between rest-frame core and total
low-frequency luminosity.

\begin{figure*}
\epsfxsize 14cm
\epsfbox{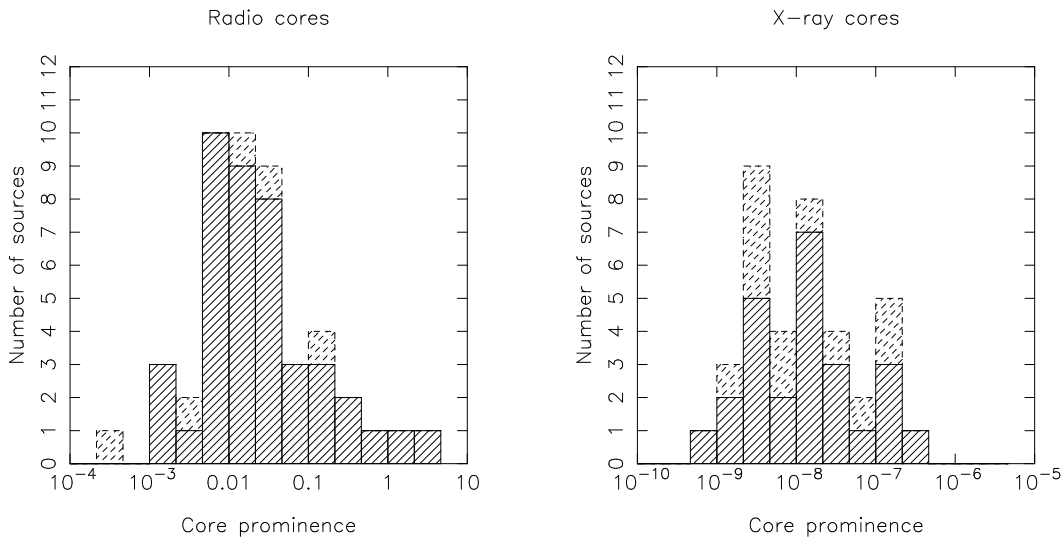}
\caption{Radio and X-ray core prominence for the B2 bright sample,
defined as discussed in the text. Dashed boxes denote upper limits on
prominence.}
\label{prom}
\end{figure*}

To use this model to generate luminosity functions, we need to
constrain $f$ and $\gamma$. We do this by examining the distribution
of radio and X-ray core prominences in the B2 sample, using the total
408-MHz luminosity as the normalizing quantity. This dispersion is
quite broad in both bands (Fig.\ \ref{prom}), so that
large values of $\gamma$ are required. We use the
approach of Hardcastle \etal\ (1999), who used the K-S test to
investigate the consistency between the probability distribution
implied by the data and that of a beaming model for a given value of
$\gamma$ and intrinsic prominence. For the Lorentz factors ($\gamma \ga 2$)
required from observations of superluminal motion in BL Lac objects,
fitting in small samples results in a degeneracy between $f$ and $\gamma$; typically, the best-fitting values
are constrained to lie along a line in $f$-$\gamma$ space which
can be shown, by integrating the Doppler beaming expression over all
angles, to be
\[
f \left\{\gamma\left[{1-{(1-1/\gamma^2)^{1/2}}\over 2}\right]\right
\}^{-(2+\alpha)} \approx \bar f_{\rm o}
\]
where $\bar f_{\rm o}$ is the mean observed prominence at the
frequency of interest. Only with samples of $\gg 100$ sources could
there be enough sources in the highly beamed
tail of the distribution to set upper and lower limits on
$\gamma$. Using the present sample, we find that the data are best fit
with $\gamma \ga 2.5$ for both the radio and X-ray. In what follows, we
choose values of $\gamma$ which obey this constraint, and
corresponding values of $f$ derived from our fits.

\subsection{Simulation}
\label{sim}

We simulated observations of BL Lac sources with sky coverage and flux
limits similar to those of the real samples. We began by generating
samples matched to the B2 sample itself; we verified that we obtained
408-MHz, 5-GHz and 1-keV luminosity functions which were consistent
with the observations (and with the analytical results of Urry \&
Schafer) and that the luminosity functions obtained were unaffected if we
extended the range of simulated luminosity beyond that observed in the
B2 sample. To simulate the optical magnitude limit, we assigned optical
absolute magnitudes to the simulated sources based on their radio
luminosities and the observed weak correlation between radio
luminosity and absolute magnitude. Next we simulated 1-Jy and EMSS
samples. The simulated 1-Jy sample was selected on the basis of radio
core flux density and did not use the optical magnitude limit of the
real sample; this is acceptable, since the optical limit in the real
data only dominates the $V_{\rm max}$ calculation in 3/34 objects. The
simulated EMSS sample was selected on X-ray core flux density and
included the effects of the variable sensitivity of the EMSS. In both
cases, we assume that all objects which meet the flux density
selection criterion are BL Lac objects, given that they have been
generated from the B2 parent population and that they are all strongly
core-dominated as a result of the selection criterion (we find that,
even for the lowest $\gamma$ values, almost all the selected objects
are more strongly core-dominated than the simulated or real B2
sources, which of course include some BL Lac objects themselves, and
that their core prominences at 5 GHz are comparable to those of real
BL Lac objects). This assumption may be an important source of error
in the simulations, and we return to it below (section
\ref{conclusion}). We also assume no evolution in the FRI parent
population, since the amount of evolution is known to be small at low
redshifts. We discuss the possible effects of evolution in section
\ref{conclusion}.

Several interesting results emerge from these simulations. Most
strikingly, the slopes of the simulated core luminosity functions for both
radio- and X-ray selected objects are systematically steeper than the
values predicted by the models of Urry \& Schafer, and more consistent
with the measured slopes (Figs \ref{5-lfbl} and \ref{1-lfbl}). It appears that
the analysis of Urry \& Schafer does not apply {\it if the beamed
sample is selected on beamed rather than unbeamed flux}; it is only
applicable if the selection flux is unbeamed (as for the B2 objects).

\subsection{Matching observations}
\label{matching}

\begin{figure*}
\epsfxsize 10cm
\epsfbox{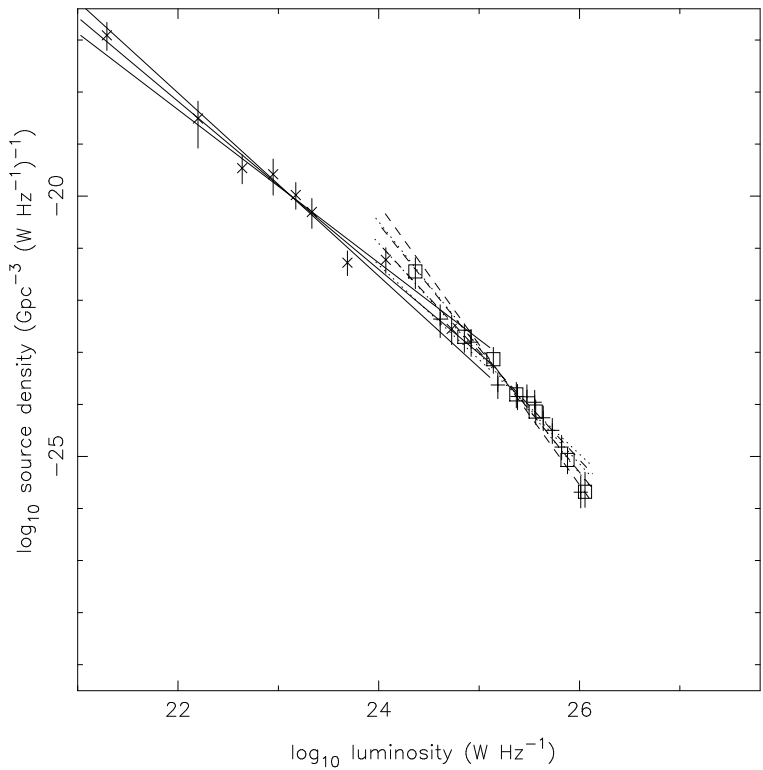}
\caption{Simulated 5-GHz core luminosity function for B2 radio galaxies,
1-Jy BL Lac objects and EMSS BL Lac objects with $\gamma = 3.5$. The
symbols and scale match those of Fig.\ \ref{5-lfbl}.}
\label{sim3}
\end{figure*}

\begin{table*}
\caption{Some results of the simulations for different model parameters}
\label{numbers}
\begin{tabular}{p{2.9cm}rrrrrrrrr}
\hline
Model&\multicolumn{3}{l}{Simulated B2 sample}&\multicolumn{3}{l}{Simulated
1-Jy sample}&\multicolumn{3}{l}{Simulated EMSS sample}\\
&Number&Slope&Intercept&Number&Slope&Intercept&Number&Slope&Intercept\\
\hline
$\gamma = 3$, no scatter&$ 55$ $( 6)$&$-1.53$ $( 0.10)$&$15.4$ $( 2.2)$&$ 21$ $( 3)$&$-2.40$ $( 0.24)$&$37.0$ $( 6.1)$&$ 32$ $( 6)$&$-2.36$ $( 0.25)$&$35.9$ $( 6.4)$\\
$\gamma = 3.5$, no scatter&$ 54$ $( 7)$&$-1.53$ $( 0.10)$&$15.3$ $( 2.3)$&$ 35$ $( 5)$&$-2.28$ $( 0.15)$&$34.1$ $( 3.9)$&$ 60$ $( 6)$&$-2.16$ $( 0.13)$&$31.0$ $( 3.3)$\\
$\gamma = 5$, no scatter&$ 57$ $( 7)$&$-1.45$ $( 0.09)$&$13.5$ $( 2.0)$&$144$ $(13)$&$-2.07$ $( 0.05)$&$29.2$ $( 1.2)$&$203$ $(14)$&$-1.99$ $( 0.06)$&$26.9$ $( 1.6)$\\[5pt]

$\gamma = 3.0$, 0.25 dex scatter&$ 55$ $( 8)$&$-1.49$ $( 0.12)$&$14.4$ $( 2.8)$&$ 29$ $( 6)$&$-2.54$ $( 0.22)$&$40.5$ $( 5.7)$&$ 42$ $( 5)$&$-2.31$ $( 0.24)$&$34.5$ $( 6.0)$\\

$\gamma = 3.0$, 0.4 dex scatter&$ 58$ $( 5)$&$-1.48$ $( 0.11)$&$14.2$ $( 2.4)$&$ 53$ $( 5)$&$-2.45$ $( 0.13)$&$38.5$ $( 3.2)$&$ 55$ $( 8)$&$-1.96$ $( 0.14)$&$25.6$ $( 3.6)$\\

$\gamma = 3.0$, 0.5 dex scatter&$ 56$ $( 6)$&$-1.45$ $( 0.13)$&$13.5$ $( 3.1)$&$ 80$ $( 9)$&$-2.38$ $( 0.06)$&$36.9$ $( 1.6)$&$ 75$ $( 8)$&$-1.88$ $( 0.20)$&$23.4$ $( 5.1)$\\[5pt]

$\gamma = 3.5$, 0.25 dex scatter&$ 56$ $( 9)$&$-1.44$ $( 0.09)$&$13.4$ $( 2.1)$&$ 51$ $( 6)$&$-2.36$ $( 0.10)$&$36.2$ $( 2.5)$&$ 78$ $( 9)$&$-2.03$ $( 0.18)$&$27.7$ $( 4.5)$\\

$\gamma = 3.5$, 0.4 dex scatter&$ 53$ $( 6)$&$-1.39$ $( 0.09)$&$12.0$ $( 2.1)$&$ 86$ $(10)$&$-2.35$ $( 0.05)$&$36.1$ $( 1.3)$&$ 98$ $( 7)$&$-1.87$ $( 0.14)$&$23.5$ $( 3.5)$\\

$\gamma = 3.5$, 0.5 dex scatter&$ 55$ $( 5)$&$-1.40$ $( 0.06)$&$12.4$
$( 1.5)$&$136$ $(12)$&$-2.29$ $( 0.05)$&$34.7$ $( 1.3)$&$124$
$(13)$&$-1.77$ $( 0.19)$&$20.8$ $( 4.7)$\\[5pt]

$\gamma = 3.5$, luminosity-dependent $\alpha_{\rm RX}$&$ 57$ $( 6)$&$-1.48$ $( 0.13)$&$14.3$ $( 2.9)$&$ 35$ $( 7)$&$-2.29$ $( 0.10)$&$34.3$ $( 2.5)$&$783$ $(27)$&$-1.79$ $( 0.03)$&$21.5$ $( 0.7)$\\

\hline
\end{tabular}
\vskip 5pt
\begin{minipage}{\linewidth}
Details of the models used are given in the text. For each model the
mean result of a number of simulations is quoted, together with the
standard deviation of that parameter in parentheses. Slopes and
intercepts are for the radio luminosity function only; the intercept
is in units of $\log_{10}($Gpc$^{-3}$ W$^{-1}$ Hz$)$. Since the B2
objects are selected on unbeamed flux, their numbers in simulated
samples are expected to be independent of beaming parameters, and are
included here for completeness only. These results are insensitive to
the choice of cosmological parameters. For comparison, the total
numbers of B2, 1-Jy and EMSS objects in the real data are 47, 34, and
41 respectively, and the slopes of their
radio core luminosity functions are $-1.68 \pm 0.16$, $-2.59 \pm
0.26$, and $-1.91 \pm 0.18$ respectively.
\end{minipage}
\end{table*}

We can now try to vary the parameters of these simulations to obtain
a better match to the observed data. There are several features of the
observed samples which we would like our simulated samples to match,
including:

\begin{enumerate}
\item the total number of sources
\item the slope and range of the luminosity functions
\item the redshift distribution, and
\item the distribution of $\alpha_{\rm RX}$.
\end{enumerate}

\begin{figure*}
\epsfxsize 18cm
\epsfbox{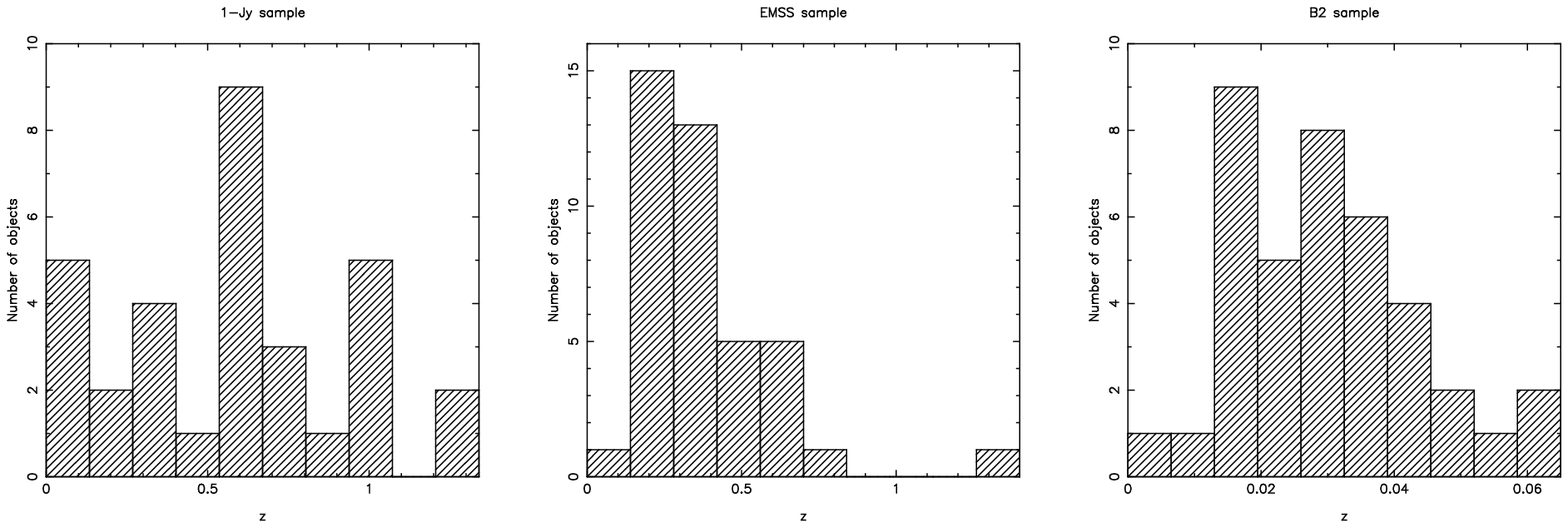}
\epsfxsize 18cm
\epsfbox{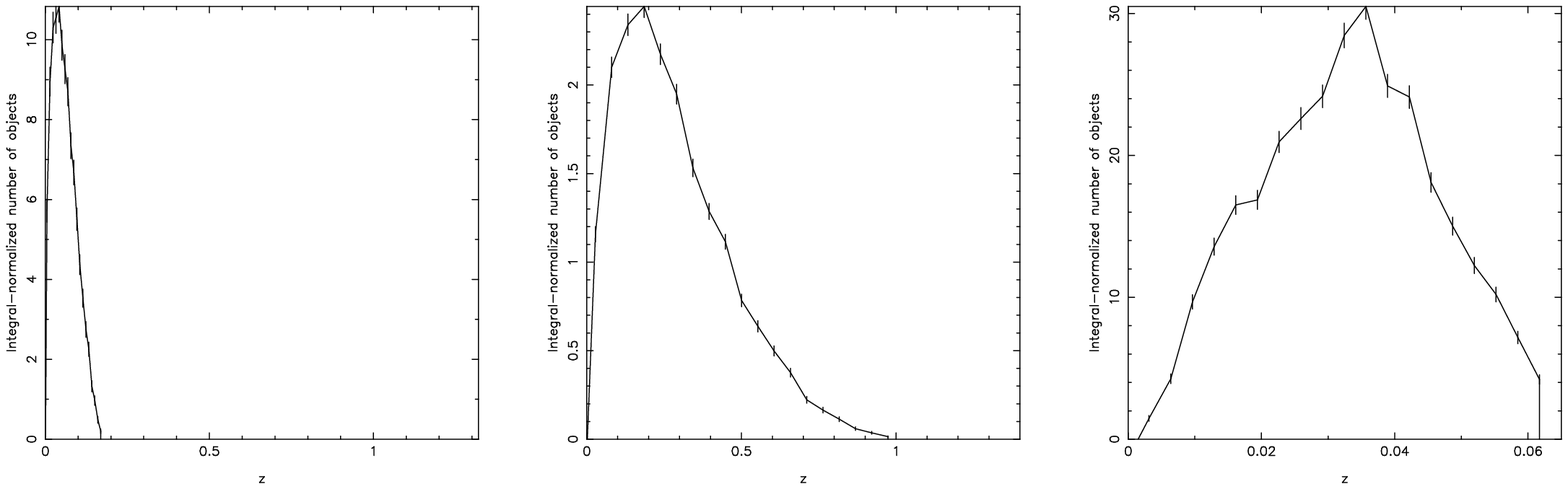}
\epsfxsize 18cm
\epsfbox{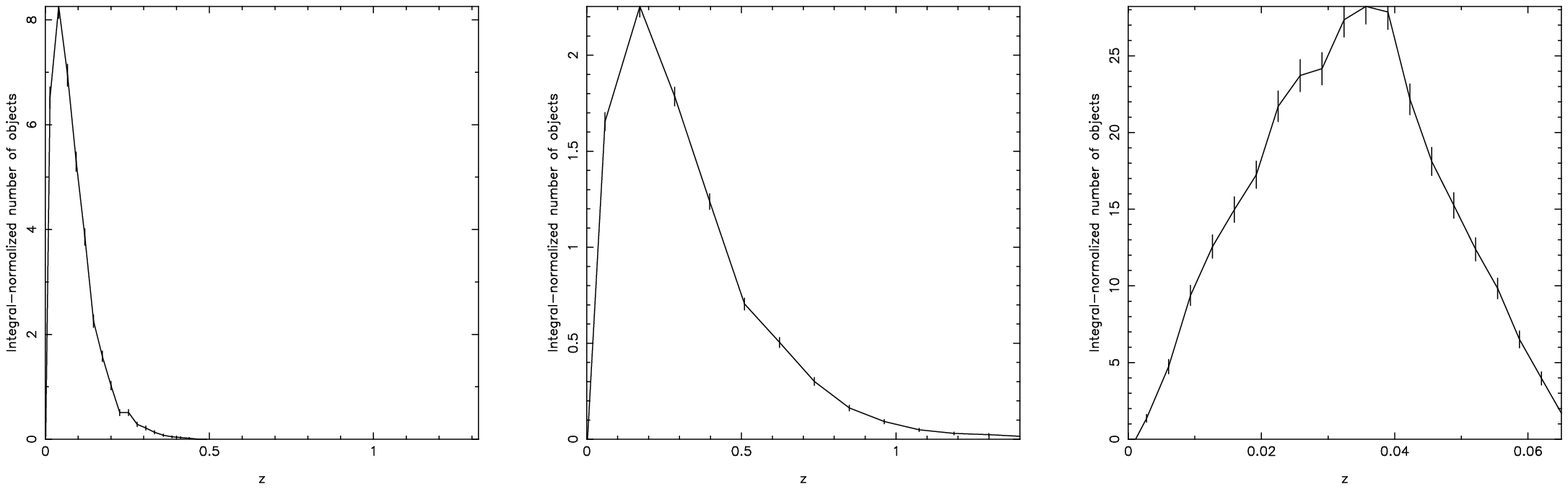}
\caption{Redshift distributions for the real and simulated samples.
Top: the real redshift distribution for the BL Lac and radio-galaxy
samples, including BL Lacs without determined redshifts, which are
assigned the median value. Middle: the distribution for simulated
samples with $\gamma = 3.5$. The EMSS sample is reasonably well
matched by the simulations, but there are no high-redshift 1-Jy
objects. Bottom: the distribution of simulated samples with $\gamma =
3.0$ and 0.4 dex of dispersion in the intrinsic radio-X-ray relation.
Simulated distributions are plotted with normalization such that the
integral under the curve is unity; see Table \ref{numbers} for the
mean number of sources in these models. Note that a much smaller
redshift range is plotted for the B2 sources than for the other objects.}
\label{redshifts}
\end{figure*}

In what follows we concentrate on reproducing the 5-GHz radio
luminosity functions, as the data in this waveband have smaller
uncertainties than the X-ray data and are less strongly affected by
factors such as variability. The inputs we are using necessarily make
the simulated B2 sample match the data on criteria (i), (ii) and
(iii), and this match is relatively insensitive to beaming parameters,
since the B2s are not selected on beamed emission. The simulated
luminosity functions for both X-ray and radio-selected BL Lac objects
in our models overlap with the luminosity function for the (simulated)
B2 objects, as in the real data. The degree to which they overlap in
luminosity, and the number of sources in the simulated BL Lac samples,
is quite a strong function of the beaming Lorentz factor $\gamma$ (and
its associated $f$ values). For $\gamma = 3$, the simulations slightly
underpredict the observed numbers of sources in both BL Lac samples
(Table \ref{numbers}), and they also predict considerably lower
minimum and maximum luminosities than are observed. With $\gamma = 5$,
we over-predict the observed numbers of sources, but do a better job
of predicting their luminosity bounds. The best results come from
simulations with intermediate values of $\gamma$, for example $\gamma
= 3.5$ (Fig. \ref{sim3}). This simulation predicts approximately the
right total number of BL Lacs in the two surveys, and it also matches
the luminosity range and the slope of the luminosity function, so that
criteria (i) and (ii) are satisfied. It does not produce enough
high-luminosity, high-redshift 1-Jy BL Lac objects, so that we cannot
claim to have satisfied criterion (iii) for this sample; the redshift
distributions of the real and simulated data are shown in Fig.\
\ref{redshifts}. However, it is known that at least some of the 1-Jy
sample are too luminous in extended emission to be FRIs and have
FRII-like radio structure; such objects cannot be generated in our
simulations. Criterion (iii) seems to be satisfied for the simulated
EMSS samples (Fig.\ \ref{redshifts}).

The numbers of BL Lacs produced in the simulations do not depend very
strongly on the adopted lower limit on the total. This tells us that
relatively few of the simulated BL Lac objects come from intrinsically
low-luminosity radio galaxies, on the reasonable assumption that these
objects do not have more strongly beamed cores than the B2s. This
lends support to the idea that the B2 radio galaxies (together, in the
case of the 1-Jy radio sources, with more luminous radio galaxies) are
indeed representative of the parent population for these samples.

\begin{figure*}
\epsfxsize 18cm
\epsfbox{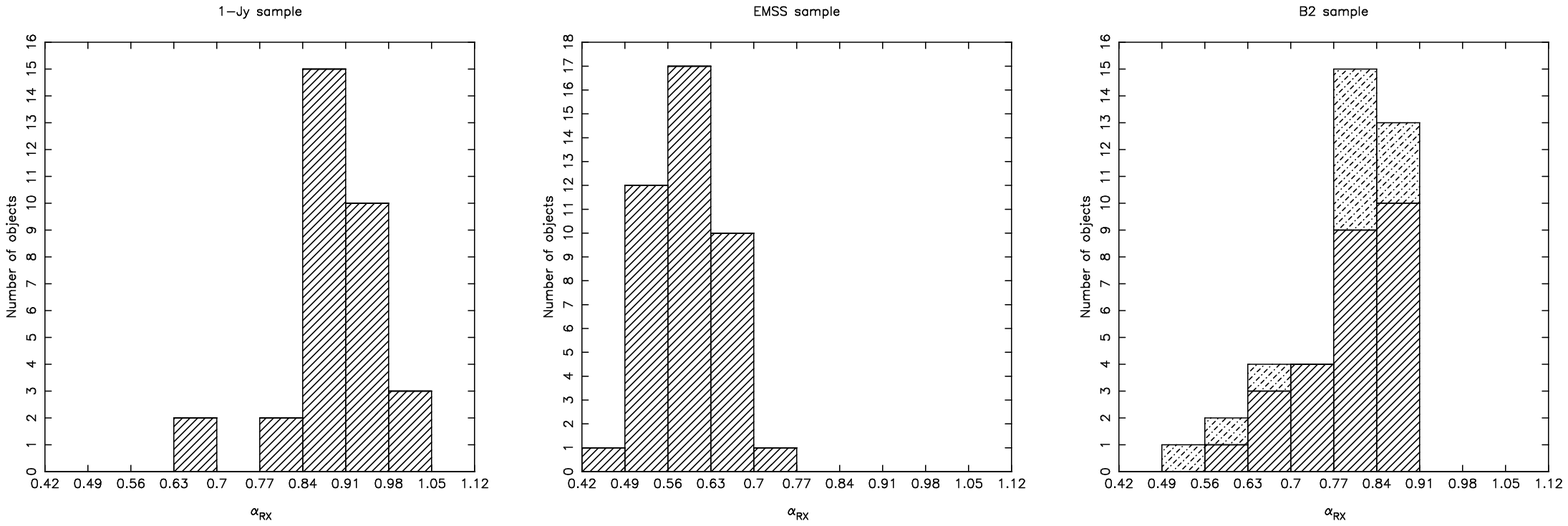}
\epsfxsize 18cm
\epsfbox{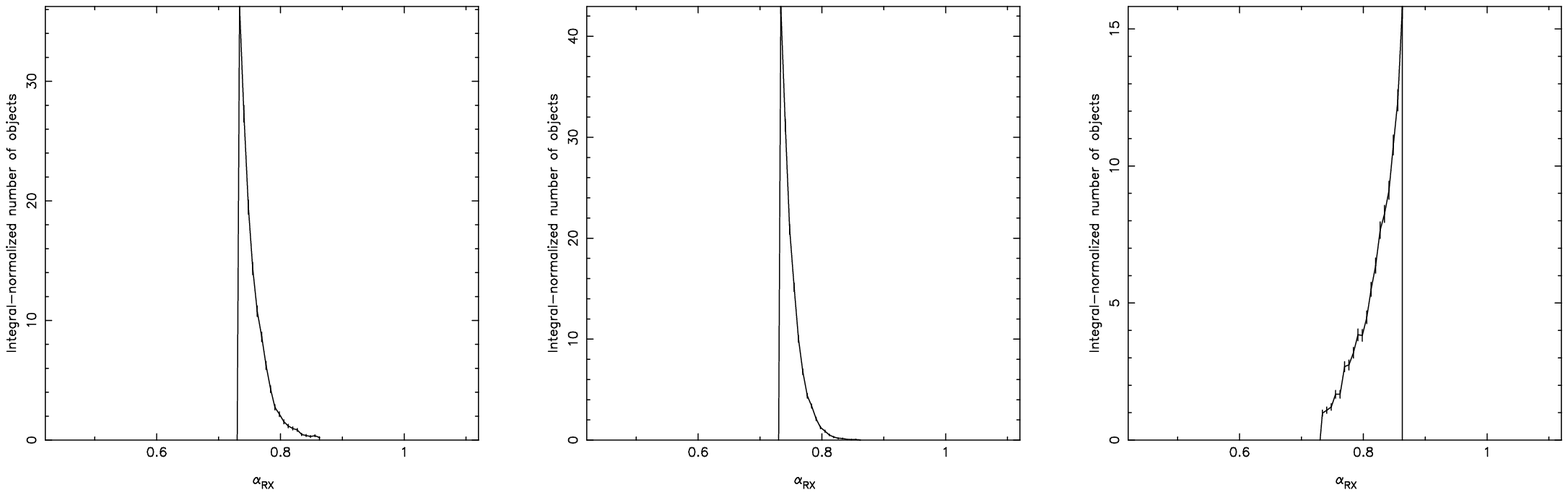}
\epsfxsize 18cm
\epsfbox{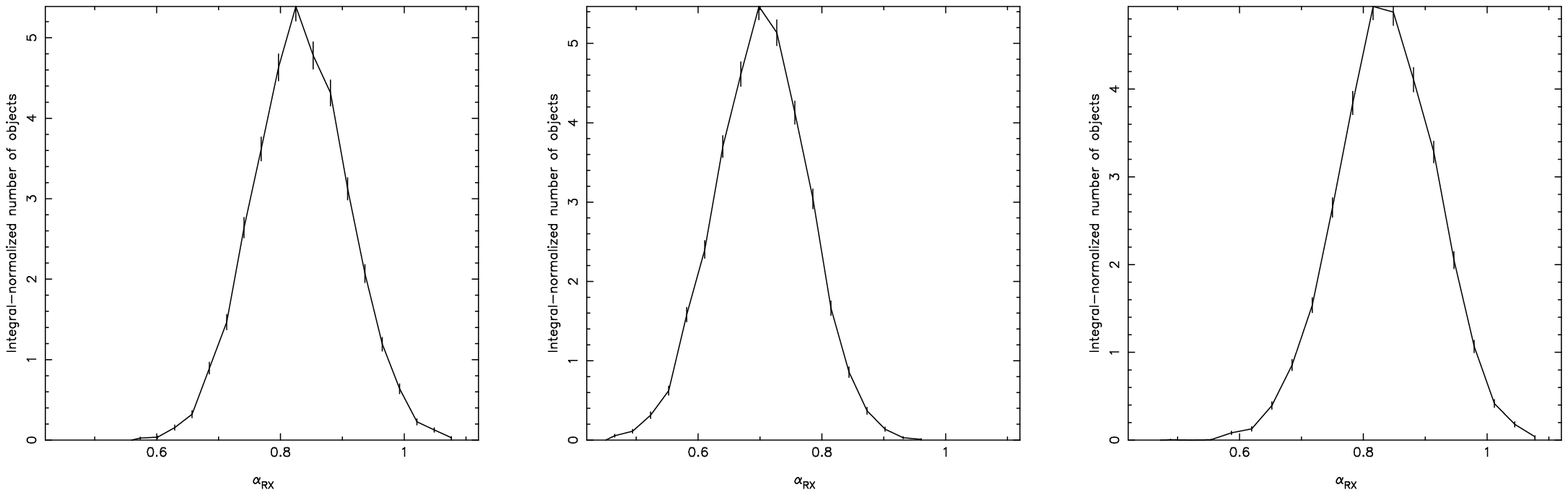}
\caption{Distribution of the radio-X-ray spectral index ($\alpha_{\rm
RX}$) for the three samples: 1-Jy BL Lac objects (left), EMSS BL Lacs
(centre) and B2 radio galaxies (right). Top: the real distribution
(dashed boxes indicate lower limits on $\alpha_{\rm RX}$). Middle: the
distribution for simulated samples with no dispersion in the
radio-X-ray relation and $\gamma=3.5$. The small dispersion in
observed $\alpha_{\rm RX}$ arises as a result of the different
spectral indices assumed for radio and X-ray bands in the Doppler
beaming calculation. The simulations clearly do not match the data.
Bottom: the distribution of simulated samples with $\gamma=3.0$ and
0.4 dex of dispersion in the rest-frame ratio of core X-ray to core
radio luminosity. This comes much closer to the observed results.
Simulated distributions are plotted with normalization such that the
integral under the curve is unity; see Table \ref{numbers} for the
mean number of sources in these models.}
\label{alphas}
\end{figure*}

These simulations fail to reproduce the observed
distinction between the luminosity functions of the X-ray and
radio-selected BL Lac objects. This is not surprising: in the
model we have considered so far, the radio and X-ray cores are both a
fixed fraction of the total low-frequency luminosity, with the result
that selection in the radio is effectively identical to selection in
the X-ray. In other words, the radio-to-X-ray spectral index of the
simulated sources, $\alpha_{\rm RX}$, is approximately fixed, which is
unrealistic. Fig.\ \ref{alphas} shows the distribution of $\alpha_{\rm
RX}$ for the real and simulated data. Criterion (iv) is not met either
for the simulated B2s or for the simulated BL Lac samples. In the next
section of the paper we discuss ways of reproducing the observed
differences in the slopes of the luminosity functions in different bands.

\begin{figure*}
\epsfxsize 10cm
\epsfbox{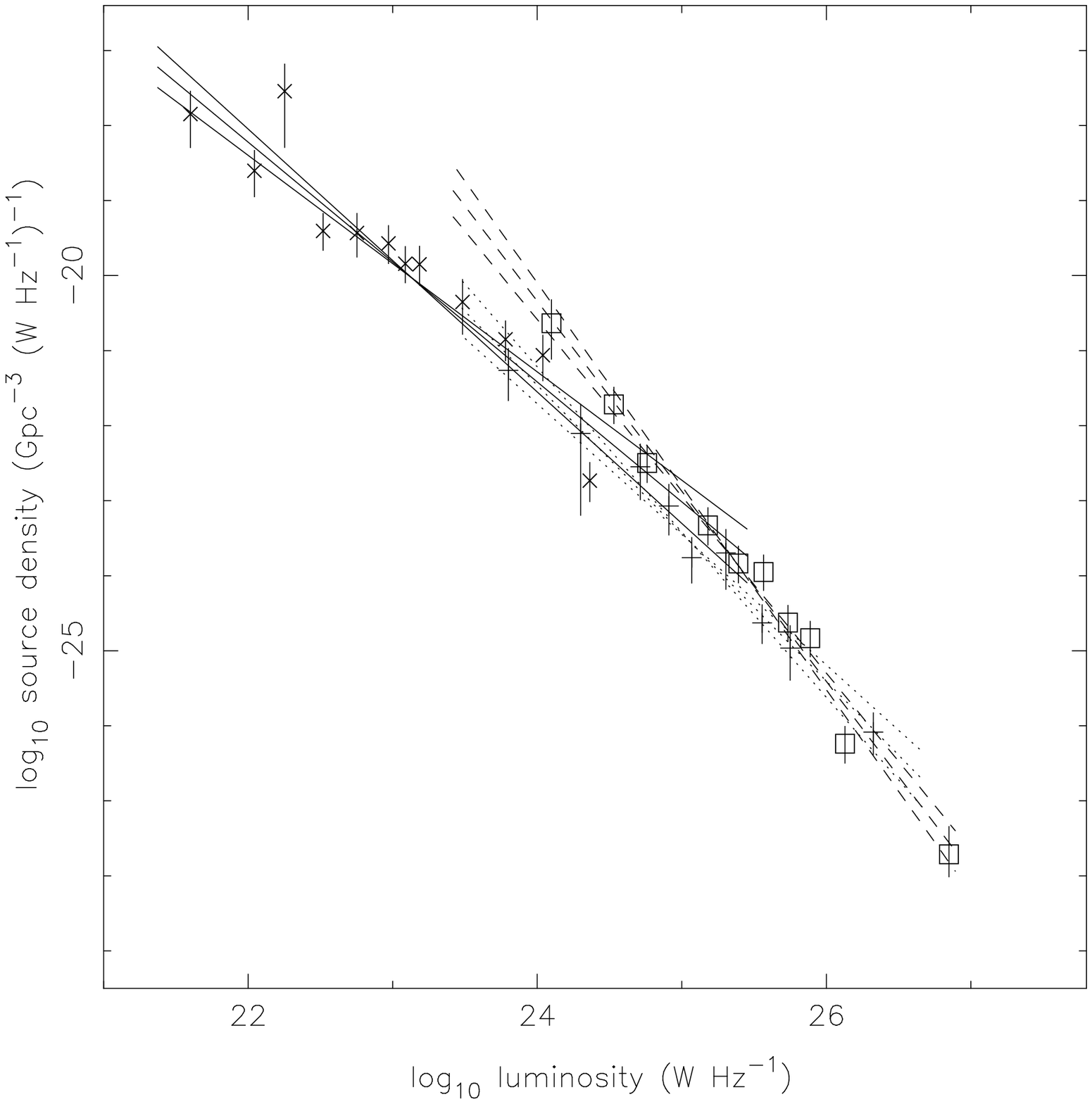}
\caption{Simulated 5-GHz luminosity function for B2 radio galaxies,
1-Jy BL Lac objects and EMSS BL Lac objects with $\gamma = 3.0$ and
logarithmic dispersion of 0.4 dex ($1\sigma$) in the intrinsic ratio
of radio and X-ray cores to extended luminosity. The symbols and scale
match those of Fig.\ \ref{5-lf}. Differences between
the luminosity functions of the two BL Lac populations are evident, as
in the real data. This model also predicts approximately the right
numbers of sources.}
\label{sim4}
\end{figure*}

\subsection{The difference between X-ray and radio-selected samples}

\subsubsection{Dispersion in the intrinsic radio-X-ray relation}
\label{dispersion}
A simple way of producing a distinction between the two BL Lac samples
is to assume that there is some dispersion, uncorrelated with
intrinsic luminosity, in the intrinsic (rest-frame) relationship
between X-ray and radio core flux, reflecting source-to-source
variation in, for example, the structure of the jet or the size and
age of the extended emission. This is distinct from, but in practice
has very similar effects to, introducing an uncorrelated dispersion in
the beaming Lorentz factors for the radio and X-ray components. Since
the most luminous sources at any given redshift are most likely to be
selected, we expect sources selected in the radio to be typically more
radio-luminous than sources in the X-ray, and vice versa. We can use
our simulations to investigate how great this dispersion should be. A
starting point is the estimated logarithmic dispersion about the
regression line in the radio-galaxy radio-X-ray core luminosity
relationship (Canosa \etal\ 1999), which was about 0.2 decades. This
is an underestimate of the dispersion in the {\it rest-frame}
relationship, because common correlation of the X-ray and radio core
luminosities with the intrinsic luminosity of the source tends to
stretch out the observed correlation. But even 0.2 dex of dispersion
begins to recover the differences between the luminosity functions of
the two BL Lac samples. Larger dispersions produce larger differences
between the slopes of the luminosity functions and the distributions
of $\alpha_{\rm RX}$, but also produce more sources (Table
\ref{numbers}). The models which come closest to satisfying criteria
(i), (ii) and (iv) above for the radio luminosity function seem to be
those which combine a low $\gamma$ with a moderate dispersion; the
luminosity functions for a representative model, with $\gamma = 3.0$
and a dispersion of 0.4 dex in the intrinsic radio-X-ray relation, are
plotted in Fig.\ \ref{sim4} (compare Fig.\ \ref{5-lfbl}) and the
distribution of spectral indices for this model is plotted in Fig.\
\ref{alphas}. These models also replicate reasonably well the slope
and range of the X-ray luminosity function for the simulated EMSS
sample. However, we are not able to reproduce the observed steep slope
of the 1-Jy X-ray luminosity function (Fig.\ \ref{1-lfbl}).

\subsubsection{Asymmetrical dispersion}
\label{asymm} 
It is of interest to ask whether an asymmetrical dispersion in the
radio-X-ray relationship can reproduce the observations in the same
way. This would correspond to the situation in which the
radio-selected objects are the tail of a population of X-ray-dominant
objects, or vice versa. We investigated this by allowing either the
radio or the X-ray intrinsic prominence to have a dispersion of 0.5
dex ($1\sigma$) above the mean values determined from the radio
galaxies; this corresponds to radio-dominance or X-ray-dominance.
Unsurprisingly, such models do not reproduce the difference between
the luminosity functions in both bands; for example, if the X-rays
have no dispersion in intrinsic prominence, then we see little
difference between the X-ray luminosity functions of radio-selected
and X-ray-selected objects, although their radio luminosity functions
are different. We need dispersion in both bands to reproduce the
differences in their luminosity functions.

\subsubsection{Different Lorentz factors}
\label{beaming}
It has been suggested (section \ref{intro}) that the differences
between X-ray and radio-selected objects can be attributed to
different Lorentz factors for the regions emitting in the two
wavebands. Our simulations show that the observed correlation between
X-ray and radio core luminosity for the B2 objects (Canosa \etal\
1999) is not consistent with isotropic X-ray emission, but we cannot
rule out models in which the X-ray nuclei are beamed with a Lorentz
factor considerably lower than that of the radio, as has been proposed
to account for the less extreme properties of X-ray selected objects.
Values of $\gamma_{\rm Xray} \la 3$ do not produce enough EMSS BL
Lacs in our model; if the Lorentz factors were this low, then these
sources would have to have a different parent population. With
$\gamma_{\rm Xray} = 3$, $\gamma_{\rm radio} = 5$ we do not see any
significant difference between the slopes or normalizations of the
populations in the two samples. Different Lorentz factors cannot alone
be responsible for the difference in the observed luminosity function
slopes and normalizations, although, by the same token, we cannot use
the observed luminosity functions to rule out mildly differing Lorentz
factors, since intrinsic dispersion in the radio-X-ray relation would
obscure their effects.

\subsubsection{Luminosity dependence of $\alpha_{\rm RX}$}
\label{ldep}
A further way to try to reproduce the difference between the radio-
and X-ray-selected BL Lac samples would be to include some luminosity
dependence into the X-ray/radio relationship, as described by Fossati
\etal\ (1998). The simplest way to parametrise this would be in terms
of an $\alpha_{RX}$ which varies with luminosity. Unfortunately, the
correlations in Fossati \etal\ are parametrised in terms of 5-GHz
luminosity, which is itself a beamed quantity; for our purposes we
need some relation between the shape of the SED and the beam power as
parametrised by the low-frequency unbeamed luminosity. To test whether
this kind of model can reproduce the observed difference in the
luminosity functions of the different samples, we made the intrinsic
X-ray flux fraction a function of intrinsic 408-MHz luminosity, while
keeping the intrinsic radio flux fraction constant. We chose a
dependence on luminosity such that $\alpha_{\rm RX}$ increases
linearly with luminosity, going from 0.6 in the least luminous sources
to 1.0 in the most luminous sources we observed; this approximately
matches the magnitude of the range in $\alpha_{\rm RX}$ seen by
Fossati \etal\ (see, e.g., their figure 8), though over a considerably
smaller range in radio luminosity. This has some unrealistic effects
-- for example, because the majority of sources are low-luminosity and
flat-spectrum, the model vastly overpredicts the number of
X-ray-selected sources (Table \ref{numbers}) and also predicts an
inconsistent $\alpha_{\rm RX}$ distribution. However, it does give
rise to clear differences between the luminosity functions of the two
simulated BL Lac samples, in the same sense as, though of smaller
magnitude than, the differences seen in the data. We conclude that
luminosity dependence of the X-ray/radio relationship is probably not
in itself enough to explain the differences in the luminosity
functions of the two populations.

\section{Summary and discussion}
\label{conclusion}

In this paper we have been trying to determine to what extent and
under what assumptions B2 radio galaxies (or, more accurately, the
radio galaxy population represented by the B2 radio galaxies) can be
the parent population of two well-studied samples of BL Lac objects
(the EMSS and 1-Jy samples). Following earlier work, we have
characterized the properties of the populations using their luminosity
functions: luminosity functions for the B2 and BL Lac samples were
constructed in sections \ref{b2lum} and \ref{bllum}. In section
\ref{usmodl} we discussed the analytical results of Urry \& Schafer
(1984); we argued that their models cannot properly be applied to
luminosity functions of samples with different selection criteria.

Instead, in sections \ref{mc} and \ref{sim}, we used Monte Carlo
simulations to set up a parent population of beamed radio galaxies
with B2-like properties and drew objects from them with selection
criteria which as closely as possible matched those actually used in
generating the EMSS and 1-Jy samples. By adjusting the unknown
parameters of the simulation (section \ref{matching}) we were able to
show that our simulated observations of a beamed population of radio
galaxies matched to the B2 sample were able to reproduce reasonably
well the observed numbers of BL Lac objects and the slopes of the
radio- and X-ray-selected BL Lac luminosity functions, if low values
($\sim 3$) of the beaming Lorentz factor $\gamma$ are adopted. In
section \ref{dispersion} we found that the introduction of a moderate
dispersion in the rest-frame ratio of core X-ray to core radio
luminosity (for which there is some direct evidence in the
observations of the B2 radio galaxies) can help to explain the
observed differences in the luminosity functions and radio-to-X-ray
spectral indices of radio-selected and X-ray-selected BL Lacs. This
approach gives better results than other proposed ways of generating
the differences between BL Lac populations (sections \ref{asymm} --
\ref{ldep}). These results therefore support a model in which the two
apparently different BL Lac populations are simply extreme objects
drawn from a single parent population (cf.\ Laurent-Muehleisen \etal\
1999), although we have not attempted to reproduce all of the observed
differences between radio-selected and X-ray-selected BL Lacs, such as
the apparent differences in cosmological evolution (section
\ref{intro}).

The beaming Lorentz factors used here are much smaller than those
inferred from superluminal motion or $\gamma$-ray transparency in some
BL Lac objects. These typically require $\gamma \ga 10$, which would
grossly overpredict the number of BL Lac objects that should be
observed above the flux limit (cf.\ Table \ref{numbers}). Chiaberge
\etal\ (2000) have previously pointed out that the
nuclear luminosities of FRI radio galaxies are a factor 10--10$^4$ too
bright to be consistent with the `de-beaming' of BL Lacs using these
high Lorentz factors. The solution they adopt, velocity structure in
the nuclear jet, seems plausible in view of what we know
about the existence of velocity-structures in the {\it
kiloparsec}-scale jets (e.g., Laing 1996). Velocity structure in the
jets should not significantly affect our general
conclusions here. If jets have velocity structure, our values of
$\gamma$ parametrize the relationship between the observed
luminosity and the angle to the line of sight, rather than
describing real physical bulk velocities. However, the details of
this parameterization make a difference to the predicted numbers of BL Lac
objects and the range in their luminosities. Although attempting to
determine a typical jet emissivity/velocity profile is beyond the
scope of this paper, it is certainly the case that luminosity
functions can be used to help to constrain it.

The simulated observations fail to reproduce the real data in two
important ways. Firstly, although the simulated EMSS sample is a good
match in redshift distribution to the observed objects (Fig.\
\ref{redshifts}, the simulated 1-Jy sample contains far too few
high-redshift objects. As discussed in section \ref{intro}, the EMSS
objects seem all to be FRI-like in their radio structure and
luminosity, while it is known that some 1-Jy objects are FRII-like;
since FRIIs have higher luminosities than the parent population we
use, it is not surprising that some of the 1-Jy sample appear at
higher redshifts. Our models predict that essentially all the 1-Jy
objects whose parent population are FRIs should have redshifts less
than $\sim 0.55$ (Fig. \ref{redshifts}), which agrees well with
the observations of Rector \& Stocke (2001). The fact that some of the
1-Jy objects may have a different parent population may also help to
explain the anomalously steep 1-Jy X-ray luminosity function.

Secondly, the simulations predict a considerably larger number of BL
Lac objects than is observed (by factors $\ga 2$) for $\gamma \ga 4$
or if a significant dispersion in intrinsic $\alpha_{\rm RX}$ is
introduced; even the models which best reproduce other
features of the population, such as the differences between the slopes
of the X-ray and radio luminosity functions, overpredict the numbers
of 1-Jy BL Lacs by a factor $\sim 1.5$. The mismatch in numbers
becomes even greater if some of the observed high-redshift 1-Jy
objects are not drawn from an FRI parent population, as discussed
above. We can attribute at least some of this effect to the optical
selection criteria applied when defining BL Lac samples. As pointed
out by March\~a \etal\ (1996), a definition in terms of the strength
of the 4000-\AA\ break measures the strength of the optical
non-thermal emission in terms of the starlight in the host galaxy,
which may have very little to do with the AGN, while a condition on
the equivalent width of the strongest emission lines has little
physical justification (and may exclude objects which are in all other
ways BL Lac objects; cf.\ Vermeulen \etal\ 1995). So some of the
predicted objects may be present in the surveys, identified as
something other than BL Lacs. However, a search in the EMSS survey for
objects intermediate between FRIs and traditional EMSS BL Lacs (Rector
\etal\ 1999) found relatively few candidates, and only a handful of
EMSS sources are directly identified with radio galaxies. It remains
possible that some of our simulated sources are identified as groups
or clusters in the EMSS survey. If, as new {\it Chandra} results
(section \ref{chandra-caveat}) suggest, we are overestimating the
radio galaxies' nuclear X-ray fluxes (section \ref{chandra-caveat})
then we will also have overestimated the intrinsic X-ray core
prominence; correcting for this would result in a (probably small)
reduction in the predicted total number of EMSS sources. In the 1-Jy
sample, there are additional optical magnitude and radio spectral
selection criteria which are not modelled in our simulations, and so
there is more scope for `hiding' the excess sources. On the other
hand, we are extrapolating our B2 luminosity function to high
redshifts to generate BL Lac objects without including the effects of
cosmological evolution of the FRI population, now reasonably
well-established (e.g., Waddington \etal\ 2001). Because there is
little evidence for evolution below $z<0.5$, where most of our BL Lac
candidates are generated (Fig. \ref{redshifts}), this does not have a
strong effect on our models, but there may be up to an order of
magnitude increase in the numbers of the most luminous sources at $z
\sim 1$. Taking this into account would lead our models to produce
$\sim 10$ additional high-redshift 1-Jy objects. Without a
detailed model for source evolution, we cannot make this more
quantitative.

Combining these factors with the large statistical
uncertainties on the predicted and actual numbers of objects, we
regard the degree of agreement between the simulations and
observations as encouraging. It supports a model in which FRI radio
galaxies in the luminosity range of the B2 bright sample are the
parent population of all the EMSS X-ray-selected BL Lac objects, and
of $\sim 50$ per cent of the radio-selected 1-Jy
objects. A higher-luminosity population, probably beamed FRIIs, must
be responsible for the remaining, higher-redshift, 1-Jy objects.

\section*{Acknowledgements}

We are grateful for support from NASA grant NAG 5-1882 and from PPARC.
CMC thanks the University of Bristol for a research studentship.

\bsp

\end{document}